# Magnetic-cations doped into a Double Perovskite Semiconductor

*Lun Jin\*, Danrui Ni, Xin Gui, Daniel B. Straus, Qiang Zhang and Robert J. Cava\**

L. Jin, D. Ni, X. Gui, D. B. Straus and R. J. Cava

Department of Chemistry, Princeton University, Princeton, NJ, 08544, USA

\* E-mails of corresponding authors: ljin@princeton.edu; rcava@princeton.edu

Q. Zhang

Neutron Scattering Division, Oak Ridge National Laboratory, Oak Ridge, TN, 37831, USA

We report two solid solutions based on magnetic ion doping of the double perovskite oxide $Sr_2GaSbO_6$: $Sr_2Ga_{1-x}Cr_xSbO_6$ ($0.1 \leq x \leq 0.4$) and $Sr_2Ga_{1-x}Fe_xSbO_6$ ($0.1 \leq x \leq 0.4$). All compositions crystallize in the same space group ($I4/m$) as their undoped parent phase $Sr_2GaSbO_6$, with the trivalent magnetic cations $Cr^{3+}$ or $Fe^{3+}$ partially substituting for non-magnetic $Ga^{3+}$ in one of the B-cation sites. The Cr- and Fe-doped phases display dominant antiferromagnetic coupling among the dopant magnetic moments, and exhibit decreasing band gaps with increasing substitution level.

# 1. Introduction

Complex metal oxides have been of enduring interest to scientists in a wide range of disciplines since they can exhibit intriguing physical properties, ranging from 'simple' magnetic and electronic states to 'complicated' collective behavior such as ferroelectricity and magnetoresistance.[1,2] Magnetic semiconductors, as a classic example of collective behavior, are materials that exhibit equally useful magnetic and semiconducting properties by utilizing the spin and charge of electrons simultaneously.[3–9] These spintronic materials are one of the keen research foci in materials science as they have the potential to significantly enhance the performance of devices for next-generation information technology.[10–13] The perovskite family, as one of the largest and most tolerant structural types in complex metal oxides, serves an important role in material designing area.[14–16] The ease of tuning the strength and sign of metal-oxygen coupling interactions by modifying the unit cell contents and crystallographic structures makes perovskites perfect candidates for future applications.

An $A_2BB'O_6$ B-site-cation-ordered double perovskite in which magnetic transition metal cations are absent, but which exhibits a band gap in the semiconducting regime, can be a perfect parent phase for fabricating dilute magnetic semiconductors. Magnetic-cation doping can then be achieved by partially replacing one of the non-magnetic B-cations with selected magnetic transition metal centers, to introduce magnetism into the semiconducting system. The preparation of cation ordered double perovskites is often not as straightforward as their cation disordered analogs, because random mixtures are often preferred unless the cation ordering is driven by strong chemical factors, such as distinguished ionic radii and/or formal oxidation states.

$Sr_2GaSbO_6$ meets all the above-mentioned criteria. Although the ionic radius of $Ga^{3+}$ (0.76 Å) is almost identical to that of $Sb^{5+}$ (0.74 Å), there are two units of charge difference between $Ga^{3+}$ and $Sb^{5+}$, which yields a complete B-site cation ordered double perovskite with rock-salt type $Ga^{3+}$-$Sb^{5+}$ ordering. To introduce magnetism into this magnetically silent system, trivalent transition metal cations $Cr^{3+}$ and $Fe^{3+}$ were selected to partially substitute for the non-magnetic B-site-cation $Ga^{3+}$, as $Sr_2CrSbO_6$[17–20] and $Sr_2FeSbO_6$[21–23] are well-established compounds. Thus, in this study, we have fabricated the undoped double perovskite parent phase $Sr_2GaSbO_6$, plus two series of doped double perovskite phases $Sr_2Ga_{1-x}Cr_xSbO_6$ ($0.1 \leq x \leq 0.4$) and $Sr_2Ga_{1-x}Fe_xSbO_6$ ($0.1 \leq x \leq 0.4$), and fully characterized them from the structural, magnetic, and optical points of view. The selected Cr and Fe dopants, while tuning the band gap in the two doped series, both lead to bulk antiferromagnetic behavior.

## 2. Experimental

Polycrystalline powder samples of the undoped parent phase $Sr_2GaSbO_6$ and its magnetic-ion doped phases $Sr_2Ga_{1-x}Cr_xSbO_6$ ($0.1 \leq x \leq 0.4$) and $Sr_2Ga_{1-x}Fe_xSbO_6$ ($0.1 \leq x \leq 0.4$) were prepared by traditional high-temperature ceramic synthesis on a 0.5 g scale. Large-scale samples (approximately 3 g) of $Sr_2GaSbO_6$ and the two 10%-doped phases $Sr_2Ga_{0.9}M_{0.1}SbO_6$ (M = Cr and Fe) were also prepared via the same approach and employed for powder neutron diffraction characterization. Suitable stoichiometric metal ratios of $SrCO_3$ (Alfa Aesar, 99.99%), $La_2O_3$ (Alfa Aesar, 99.99%, dried at 900 °C), $Ga_2O_3$ (Alfa Aesar, 99.999%), $Sb_2O_5$ (Alfa Aesar, 99.998%), $Cr_2O_3$ (Alfa Aesar, 99.97%) and $Fe_3O_4$ (Alfa Aesar, 99.997%) were thoroughly ground together by using an agate mortar and pestle, and then transferred into an alumina crucible for calcination. These reaction mixtures were first heated to 1000 °C in air at a ramp rate of 1 °C/min and held overnight to decompose the carbonate, and then were annealed in air at 1300 – 1400 °C (ramp rate = 3 °C/min, reaction temperature varies slightly with the doping level) for 3 periods of 72 hours with intermittent grindings between heating periods. Routine inspections were carried out by laboratory X-ray powder diffraction data collected at room temperature on a Bruker D8 FOCUS diffractometer (Cu Kα) over a 2θ range between 5° and 70° after each heating period. Once the target product phases are pure and homogeneous, laboratory XRD data with much better statistical significance, covering a 2θ range between 5° and 110°, were collected from each sample. Room-temperature time-of-flight (TOF) neutron powder diffraction data were collected at Oak Ridge National Laboratory's Spallation Neutron Source, POWGEN beamline, using neutron beam with a center wavelength of 0.8 Å. Detailed structural information analysis was performed by the Rietveld method[24] using the GSAS-II program. The particle morphology of selected compositions was examined by using an FEI XL30 field-emission gun scanning electron microscope (SEM) equipped with an Oxford X-Max 20 energy-dispersive X-ray spectroscope (EDX) running on AZtec software.

The magnetization data were collected using the Quantum Design Physical Property Measurement System (PPMS) equipped with a vibrating sample magnetometer (VSM) accessory. Temperature-dependent magnetization ($M$) data were collected from finely ground powders under an applied external field ($H$) of 1000 Oe. The resulting magnetic susceptibility χ (defined as $M$ (in emu)/$H$ (in Oe)) was then plotted against temperature. Field-dependent isothermal magnetization data between $H$ = 90000 Oe and –90000 Oe were collected at $T$ = 300 K and 2 K and plotted against field.

The diffuse reflectance spectra were collected from the powder sample of each composition at ambient temperature on a Cary 5000i UV-VIS-NIR spectrometer equipped with an internal DRA-2500 integrating sphere. The data were converted from reflectance to pseudo absorbance via applying the Kubelka–Munk function, and band gaps were calculated from Tauc plots by using both direct and indirect equations.[25]

The band structures and electronic densities of states (DOS) for were calculated for $Sr_2GaSbO_6$, $Sr_2CrSbO_6$ and $Sr_2FeSbO_6$ using the WIEN2k program. The full-potential linearized augmented plane wave (FP-LAPW) method with local orbitals was used.[26,27] The generalized gradient approximation was used for dealing with electron correlation.[28] Reciprocal space integrations were completed over a 6×6×4 Monkhorst-Pack $k$-point mesh.[29] Spin-orbit coupling (SOC) effects were only applied for the Sb atom. Spin-polarization was only employed for the Cr and Fe atoms. The structural lattice parameters of $Sr_2GaSbO_6$[30], $Sr_2CrSbO_6$[18] and $Sr_2FeSbO_6$[21] were obtained from the literature. With these settings, the calculated total energy converged to less than 0.1 meV per atom.

## 3. Results and Discussion

### 3.1 Structural Characterization

$Sr_2GaSbO_6$ is a B-site cation ordered double perovskite consisting of alternating $GaO_6$ and $SbO_6$ octahedra in the rock-salt-type ordering scheme. Both standard laboratory X-ray powder diffraction (XRD) data and TOF neutron powder diffraction data (NPD) collected from the fine grey $Sr_2GaSbO_6$ powder can be indexed by a body centered tetragonal unit cell, with lattice parameters a = 5.54090(11) Å, c = 7.90490(9) Å, which is consistent with the data found in the literature[30]. Observed, calculated and difference plots from the Rietveld refinement of $Sr_2GaSbO_6$ (space group $I4/m$) against the neutron powder diffraction data are shown in the Supporting Information (SI) Figure S1, with the detailed structural information listed in Table S1.

Since $Sr_2GaSbO_6$ is a magnetically silent semiconductor, and thus substitution of the non-magnetic B-site cation $Ga^{3+}$ by the trivalent magnetic cations $Cr^{3+}$ and $Fe^{3+}$ should introduce magnetism into the system. The success of the doping of the double perovskite host was illustrated not only by the color changes of the doped samples (pink Cr-doped sample and yellow Fe-doped sample) compared to the grey color of the undoped parent $Sr_2GaSbO_6$ phase, but also by the shift of peaks in the lab XRD patterns. The detailed structural characterization of these two 10%-doped phases $Sr_2Ga_{0.9}M_{0.1}SbO_6$ (M = Cr /Fe) was performed by refining the time of flight (TOF) neutron powder diffraction data collected at 300 K against the reported structural model of their parent phase $Sr_2GaSbO_6$[30], with the all-$Ga^{3+}$ site replaced by a mixture of 90% $Ga^{3+}$ and 10% $Cr^{3+}$/$Fe^{3+}$. These two refinements both converged smoothly and provided satisfactory agreement parameters (Cr-doped: $wR$ = 5.687%, $GOF$ = 4.21; Fe-doped: $wR$ = 5.762%, $GOF$ = 4.03). Observed, calculated and difference plots from the Rietveld refinements of $Sr_2Ga_{0.9}M_{0.1}SbO_6$ (M = Cr/Fe, space group $I4/m$) against neutron powder diffraction data collected at 300 K are shown in Figure 1a & b, with the detailed structural information presented in Table 1 & 2. The structural model of $Sr_2Ga_{0.9}M_{0.1}SbO_6$ (M = Cr/Fe), and the bonding environments within the polyhedra are depicted in Figure 1c. Selected bond lengths for these polyhedra are listed in Table S2.

The attempts of 10% Cr-for-Ga and Fe-for-Ga substitution while the B-site-cation-ordering is still conserved indicates that $Sr_2GaSbO_6$ is capable of forming a partial solid-solution with $Sr_2CrSbO_6$ or $Sr_2FeSbO_6$ while maintaining its rigorous B-site cation ordering to a certain extent. Thus investigations of the remainder of the phase diagrams of these two solid-solutions were carried out by the fabrication and testing of materials with 10% increments in doping levels. The Ga-Cr and Ga-Fe systems both exhibit an upper doping limit of around 40% M-for-Ga substitution in a tetragonal symmetry ordered double perovskite phase. For higher levels of doping the materials become cation-disordered perovskite solid solutions. The presence of the disordered phase is evidenced for example by the appearance of a pseudo-cubic peak in the middle of the tetragonal doublet around the 2θ position of 46° in the laboratory XRD patterns for $Sr_2Ga_{0.6}M_{0.4}SbO_6$ (M = Cr/Fe), marked by the asterisk in Figure 2a. Thus, for compositions $x \geq 0.4$ in the $Sr_2Ga_{1-x}M_xSbO_6$ series (M = Cr/Fe), the samples are not single pure phase representatives of the B-site cation ordered double perovskite - but also contain B-site cation disordered single perovskites. These disordered phases are not the subject of the current work.

To further illustrate the successful synthesis of the doped phases, lattice parameters were determined by Rietveld refinements of the laboratory XRD data, hence the resulting lattice parameters $a$, $c$ and cell volume $V$ are plotted against the doping level $x$ in $Sr_2Ga_{1-x}M_xSbO_6$ for each dopant in Figure 2b. They exhibit significant curvature with respect to the doping level $x$ for both Cr and Fe series. Lattice parameters and cell volumes of both end-members $Sr_2CrSbO_6$[20] and $Sr_2FeSbO_6$[22] are extracted from the literature (data points in red circles) and all fall in line with the materials reported in this study.

Finally, scanning electron microscope (SEM) images (Figure S2) were obtained from selected compositions in the $Sr_2Ga_{1-x}M_xSbO_6$ series (M = Cr/Fe) to observe and compare the influence on particle morphologies brought by dopant contents. The particle morphologies unambiguously show a decrease in grain size upon elevating the doping level. Energy-dispersive X-ray spectroscopic (EDX) analysis provided compositional results that coincide with the nominal ones within the experimental uncertainties, again proving that the doping was successful.

### 3.2 Magnetic Characterization

The temperature-dependent and field-dependent magnetization data collected for each composition of the $Sr_2Ga_{1-x}M_xSbO_6$ series (M = Cr/Fe) are plotted as magnetic susceptibility $\chi$ ($M/H$) against temperature $T$ and magnetic moment $M$ against field $H$, respectively, in the Supporting Information (SI) Figures S3 – S10. In Figure 3, in order to directly analyze the influences on magnetic properties exerted by dopants within these two doped series, the magnetic susceptibility $\chi$ ($M/H$), corrected by a small temperature-independent term $\chi_0$, is plotted against temperature $T$, hence the first derivatives are summarized in the embedded plot to show the deviation from typical Curie-Weiss behavior; and the magnetic moment $M$ collected at 2 K is plotted against field $H$ as well for each composition of the Cr- and Fe-doped $Sr_2Ga_{1-x}M_xSbO_6$ series. The stacking profiles of 2 K isotherms clearly indicate that the system becomes less magnetically saturated with higher contents of dopants in both doped series. The magnetic susceptibility data, over the suitable temperature range (selected as the straight-line part of the $1/\chi$ vs T curves) of these doped phases were then fitted to the Curie-Weiss law ($\chi = C/(T - \theta) + \chi_0$), in which a set of values of Curie constant $C$ and Weiss temperature $\theta$ that are extracted and listed in Table S3 and depicted in Figure 4a & b. The magnetization data for both end-members $Sr_2CrSbO_6$[18] and $Sr_2FeSbO_6$[20] are extracted from the literature (data points in red circles). Those data all fall in line with the materials reported in this study.

For all four Cr-doped double perovskite materials, $Sr_2Ga_{1-x}Cr_xSbO_6$ ($0.1 \leq x \leq 0.4$), the magnetic susceptibility data in the whole measured temperature range $1.8 \leq T / K \leq 300$ can be fitted to the Curie-Weiss law. The inverse of $\chi$ exhibits a perfect linear relationship with temperature $T$ and the data extracted agrees with the frequently observed high-spin $d^3$ electronic configuration of $Cr^{3+}$. The Weiss temperature $\theta$ progresses from an approximately-zero value to a small negative value, which suggests that a weak antiferromagnetic interaction dominates upon elevating the doping level of $Cr^{3+}$ (Figure 4). The isothermal magnetization data collected from the Cr-doped materials at 300 K as a function of applied field is linear and passes through the origin, with a small positive slope, while the analogous data collected at 2 K exhibit some 'S-shape' features with positive slopes as well, and does not exhibit any hysteresis, suggesting canted antiferromagnetic behavior, consistent with the temperature-dependent magnetization data. The 'S-shape' features in the 2 K isotherms become less significant and progressively approach a linear shape with increasing substitution level $x$ (Figure 3).

The four Fe-doped double perovskites, $Sr_2Ga_{1-x}Fe_xSbO_6$ ($0.1 \leq x \leq 0.4$) exhibit similar bulk magnetic properties compared to their Cr analogs, but the inverse of $\chi$ exhibits a perfect linear relationship with temperature $T$ within a smaller temperature range ($30 \leq T / K \leq 300$). The data extracted are consistent with a high-spin $d^5$ electronic configuration for $Fe^{3+}$. The Weiss temperatures $\theta$ extracted are all negative and increase in magnitude upon increasing the doping level of $Fe^{3+}$, which suggests that antiferromagnetic interactions become much stronger with increasing $x$ for $Sr_2Ga_{1-x}Fe_xSbO_6$ (Figure 4). The isothermal magnetization data collected from the Fe-doped phases at 300 K is linear as a function of applied field and passes through the origin, with a small positive slope; analogous data collected at 2 K have positive slopes as well and do not exhibit any hysteresis. The much more prominent 'S-shape' features observed in the 2 K isotherms of the Fe-doped phases compared to their Cr-doped analogs are evidence for canted antiferromagnetic behavior, consistent with the temperature-dependent magnetization data. Just like their Cr analogs, the Fe-doped double perovskite also progressively approaches a linear M vs H curve for increasing doping level $x$ (Figure 3).

In Figure 4c, the observed effective moment per formula unit $\mu_{eff.obs}$ is plotted against the calculated effective moment per formula unit $\mu_{eff.cal}$ extracted based on the spin-only value of the transition metal magnetic moment. The trend lines align perfectly along the diagonal of the quadrant for both Cr and Fe cases, which means that the observed $\mu_{eff.obs}$ and theoretical $\mu_{eff.cal}$ completely coincide. This reflects the presence of neglectable orbital angular momentum contributions for magnetism in these two series. In Figure 4d, the observed magnetization M per formula unit collected at $T = 2$ K under an applied field of 9 T is plotted against the calculated saturation magnetization $M_s$ per formula unit. For the Fe-doped double perovskite series, the observed magnetization M is almost independent of the Fe-doping concentration, and thus the trend line is nearly parallel to the *x*-axis. As a clear distinction from the Fe analogs, in the Cr-doped series the observed magnetization M increases with elevated doping level *x*, but the increments gradually reduce with respect to the 10% intervals in dopant concentrations.

### 3.3 Band Structures and Band Gaps

The band structures of the undoped parent phase $Sr_2GaSbO_6$ and two end-members $Sr_2CrSbO_6$ and $Sr_2FeSbO_6$ are calculated by the method described in the experimental section. For the undoped double perovskite $Sr_2GaSbO_6$ (Figure 5a), a direct bandgap is exhibited at the $\Gamma$ point (~ 1 eV) while a much larger indirect one (~3 eV) is observed between the $\Gamma$ and N points. A nearly flat band which sits almost on top of the Fermi level is calculated to exist between the $\Gamma$ and X points. For both end-members $Sr_2CrSbO_6$ (Figure 5b) and $Sr_2FeSbO_6$ (Figure 5c), the spin structures for the calculations were set to be antiferromagnetic based on the data found in the literature[18,21]. Firstly, both calculations suggest that both $Sr_2CrSbO_6$ and $Sr_2FeSbO_6$ have significantly smaller bandgaps than non-magnetic $Sr_2GaSbO_6$. A direct small, calculated bandgap (~0.2 eV) emerges at the S point for $Sr_2CrSbO_6$, while a flat continuous bandgap (~0.1 eV) is observed between the $\Gamma$ and Z points for $Sr_2FeSbO_6$. Secondly, $Sr_2CrSbO_6$ has a nearly flat band just below the Fermi level through the Brillouin zone, while $Sr_2FeSbO_6$ has one just above. Thus based on the calculation, electron-doping of $Sr_2FeSbO_6$, to push the Fermi level closer to the energy of the flat band may be of future interest.

To experimentally evaluate the bandgaps of the $Sr_2Ga_{1-x}M_xSbO_6$ series (M = Cr/Fe) prepared in this study, the diffuse reflectance spectra were collected from powder samples at ambient temperature. The pseudo-absorbance (transferred from the observed reflectance using the Kubelka-Munk function) is plotted against photon energy (eV) in Figure 6a & b. The optical band gaps of these phases were then analyzed by Tauc plots using both direct and indirect transition equations (Table S4). The band gap of the undoped double perovskite $Sr_2GaSbO_6$ host is calculated to be 3.83 eV via the direct-transition approach, while it is found to be 3.52 eV via the indirect one. Thus the pale grey appearance of the prepared $Sr_2GaSbO_6$ polycrystalline powders clearly agrees more with the experimental bandgaps obtained from the diffuse reflectance spectra than the calculated ones obtained from the band structure calculations (from both the literature[31] and our own). It is not surprising as DFT calculations commonly underestimate band gaps for compounds containing main group elements.

For consistency, the largest transition in each phase was taken to experimentally evaluate the resulting bandgap, and we found that in general the size of band gap decreases with increasing doping level $x$ in both Cr- and Fe-doped systems, consistent with the samples' darker appearances, and though generally similar, different-in-detail behavior is observed in the optical data for each doped series (Figure 6). The Cr- and Fe- doped phases have similar absorption behavior in the 3 to 4 eV range but are differentiated with some weaker absorption tails or jumps in the lower energy regime. This can be a consequence of some strongly absorbing localized energy states in the band gap or $d$-$d$ transitions of the transition metal dopants, which have been previously observed in other polycrystalline semiconductor materials[32,33]. This is also consistent with the different appearances of the Cr- and Fe- doped phases, i.e., Cr-doped ones clearly have stronger absorption tails than Fe-doped ones (Figure 6a & b), and indeed the powders of the Cr-doped materials display a darker color compared to the Fe-doped ones. The optical data for the Fe-doped material seems to evolve very systematically to lower values, with no separate optical absorption. The band gaps plotted in Figure 6c are all calculated from Tauc plots by using an indirect transition equation;

these values will be larger if a direct transition equation is used instead (Table S4) although the trend remains the same. The only anomaly that does not fall in line in either series is $Sr_2Ga_{0.6}Cr_{0.4}SbO_6$, it has a larger bandgap than can be expected from a simple atomic substitution. Thus using either the direct or indirect approach, the band gaps measured by diffuse reflectance spectroscopy serve as strong evidence that the Cr- and Fe-doped $Sr_2Ga_{1-x}M_xSbO_6$ double perovskites are semiconductors with tunable band gaps.

## 4. Conclusions

The B-site cation ordered double perovskite $Sr_2GaSbO_6$ and its magnetic-cation doped phases $Sr_2Ga_{1-x}Cr_xSbO_6$ ($0.1 \leq x \leq 0.4$), $Sr_2Ga_{1-x}Fe_xSbO_6$ ($0.1 \leq x \leq 0.4$) were synthesized by a conventional solid state reaction method and studied from structural, magnetic and optical points of view. The doped materials crystallize in the same space group ($I4/m$) as their undoped parent phase $Sr_2GaSbO_6$, with the $Ga^{3+}/M^{3+}$ cations completely ordered with $Sb^{5+}$. The upper doping limit was found to be 40% in both systems, as evidenced by the appearance of a pseudo-cubic peak in the middle of the tetragonal doublet around the 2θ position of 46° in the laboratory XRD patterns, which represents the growth of B-site cation disordered perovskites. Magnetization data collected from the doped phases show that the Cr- and Fe-doped materials exhibit antiferromagnetic interactions, which become stronger with increasing doping levels. The diffuse reflectance spectra collected indicate that the bandgap of the undoped parent phase $Sr_2GaSbO_6$ has been reduced upon Cr- or Fe-doping, and in general the bandgap shows a decreasing trend with respect to the increasing doping level. It is not surprising, since according to our band structure calculations, both end members $Sr_2CrSbO_6$ and $Sr_2FeSbO_6$ have significant smaller bandgaps than $Sr_2GaSbO_6$. For comparison, as reported in one of our other studies, Mn dopants can introduce tunable ferromagnetism and bandgaps into the magnetically silent oxide double perovskite semiconductor as well.[34] Therefore, by selecting the appropriate magnetic dopant and controlling the doping level, magnetic semiconductors with desired physical properties can be fabricated.

**Supporting Information**

The Supporting Information is available from the Wiley Online Library or from the author.

**Acknowledgements**

This research was primarily done at Princeton University, supported by the US Department of Energy, Division of Basic Energy Sciences, grant number DE-FG02-98ER45706. A portion of this research used resources at the Spallation Neutron Source, a DOE Office of Science User Facility operated by the Oak Ridge National Laboratory. The authors acknowledge the use of Princeton's Imaging and Analysis Center, which is partially supported through the Princeton Center for Complex Materials (PCCM), a National Science Foundation (NSF)-MRSEC program (DMR-2011750).

**Conflict of Interest**

The authors declare no conflict of interest.

**Table 1.** Structural parameters and crystallographic positions from the refinement of neutron powder diffraction data collected from $Sr_2Ga_{0.9}Cr_{0.1}SbO_6$ at 300 K.

| Atoms | $x/a$ | $y/b$ | $z/c$ | S.O.F. | $U_{iso}$ equiv. ($Å^2$) |
|---|---|---|---|---|---|
| Sr1 | 0 | 0.5 | 0.25 | 1 | 0.00611 |
| Ga1 | 0 | 0 | 0 | 0.9 | 0.00126 |
| Sb1 | 0.5 | 0.5 | 0 | 1 | 0.00339 |
| O1 | 0 | 0 | 0.2495(2) | 1 | 0.01026 |
| O2 | 0.2245(2) | 0.2759(5) | 0 | 1 | 0.00733 |
| Cr1 | 0 | 0 | 0 | 0.1 | 0.00126 |

$Sr_2Ga_{0.9}Cr_{0.1}SbO_6$ space group $I4/m$ (#87)

Formula weight: 460.95 g mol$^{-1}$, Z = 2

$a$ = 5.5446(9) Å, $c$ = 7.9044(1) Å, Volume = 243.010(4) Å$^3$

Radiation source: time of flight neutrons

Temperature: 300 K

$wR$ = 5.687%; $GOF$ = 4.21

**Table 2.** Structural parameters and crystallographic positions from the refinement of neutron powder diffraction data collected from $Sr_2Ga_{0.9}Fe_{0.1}SbO_6$ at 300 K.

| Atoms | $x/a$ | $y/b$ | $z/c$ | S.O.F. | $U_{iso}$ equiv. ($Å^2$) |
|---|---|---|---|---|---|
| Sr1 | 0 | 0.5 | 0.25 | 1 | 0.00619 |
| Ga1 | 0 | 0 | 0 | 0.9 | 0.00152 |
| Sb1 | 0.5 | 0.5 | 0 | 1 | 0.00330 |
| O1 | 0 | 0 | 0.2489(7) | 1 | 0.00986 |
| O2 | 0.2238(3) | 0.2767(5) | 0 | 1 | 0.00750 |
| Fe1 | 0 | 0 | 0 | 0.1 | 0.00152 |

$Sr_2Ga_{0.9}Fe_{0.1}SbO_6$ space group $I4/m$ (#87)

Formula weight: 461.33 g mol$^{-1}$, Z = 2

$a$ = 5.5463(1) Å, $c$ = 7.9116(5) Å, Volume = 243.373(3) Å$^3$

Radiation source: time of flight neutrons

Temperature: 300 K

$wR$ = 5.762%; $GOF$ = 4.03

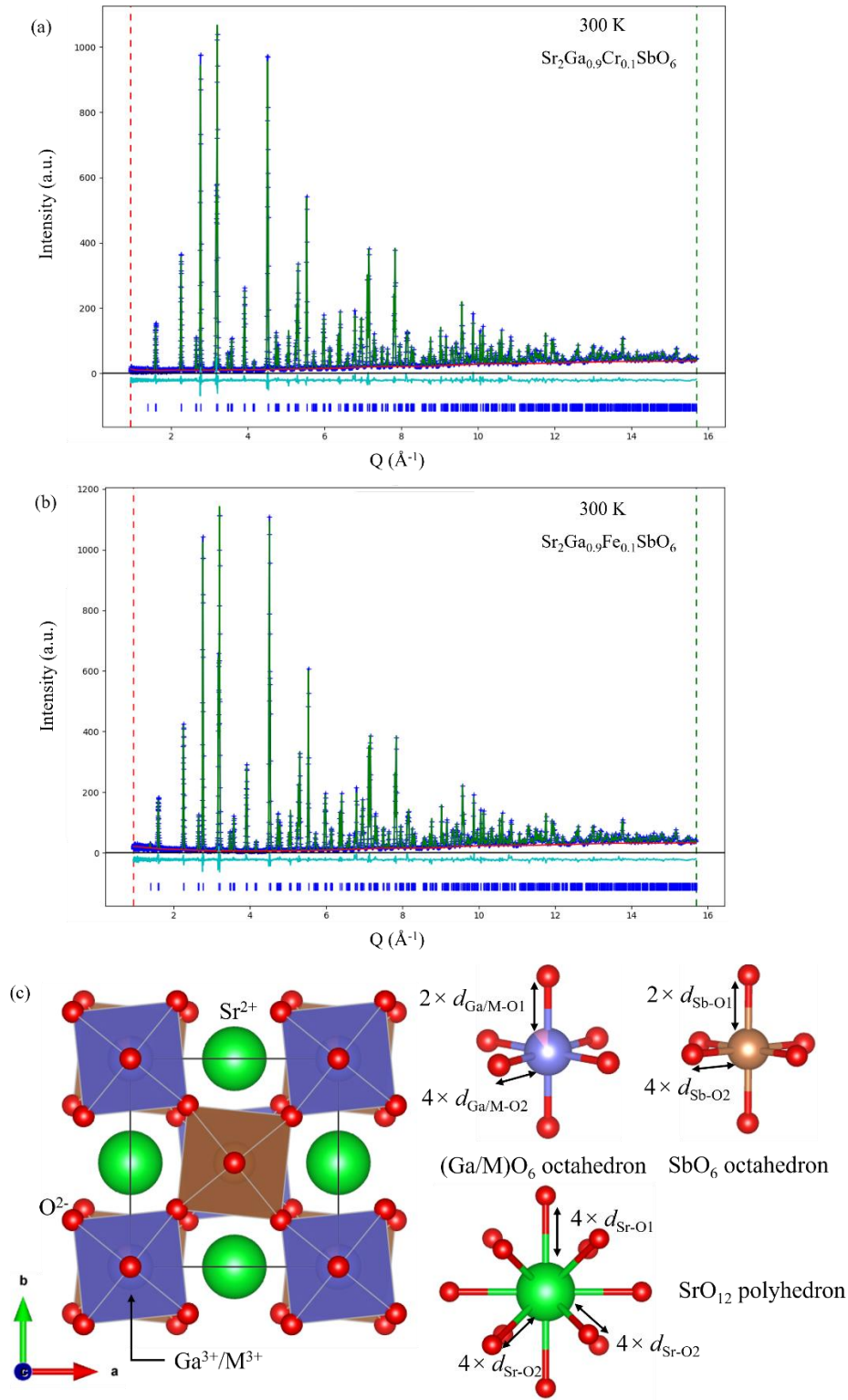

**Figure 1.** Observed, calculated and difference plots from the Rietveld refinement of (a) $Sr_2Ga_{0.9}Cr_{0.1}SbO_6$ and (b) $Sr_2Ga_{0.9}Fe_{0.1}SbO_6$ (space group $I4/m$) against neutron powder diffraction data collected at 300 K; (c) The structural model of $Sr_2Ga_{0.9}M_{0.1}SbO_6$ (M = Cr/Fe), and the selected bonding environment of the $Ga_{0.9}M_{0.1}O_6$ octahedron, $SbO_6$ octahedron and the $SrO_{12}$ polyhedron.

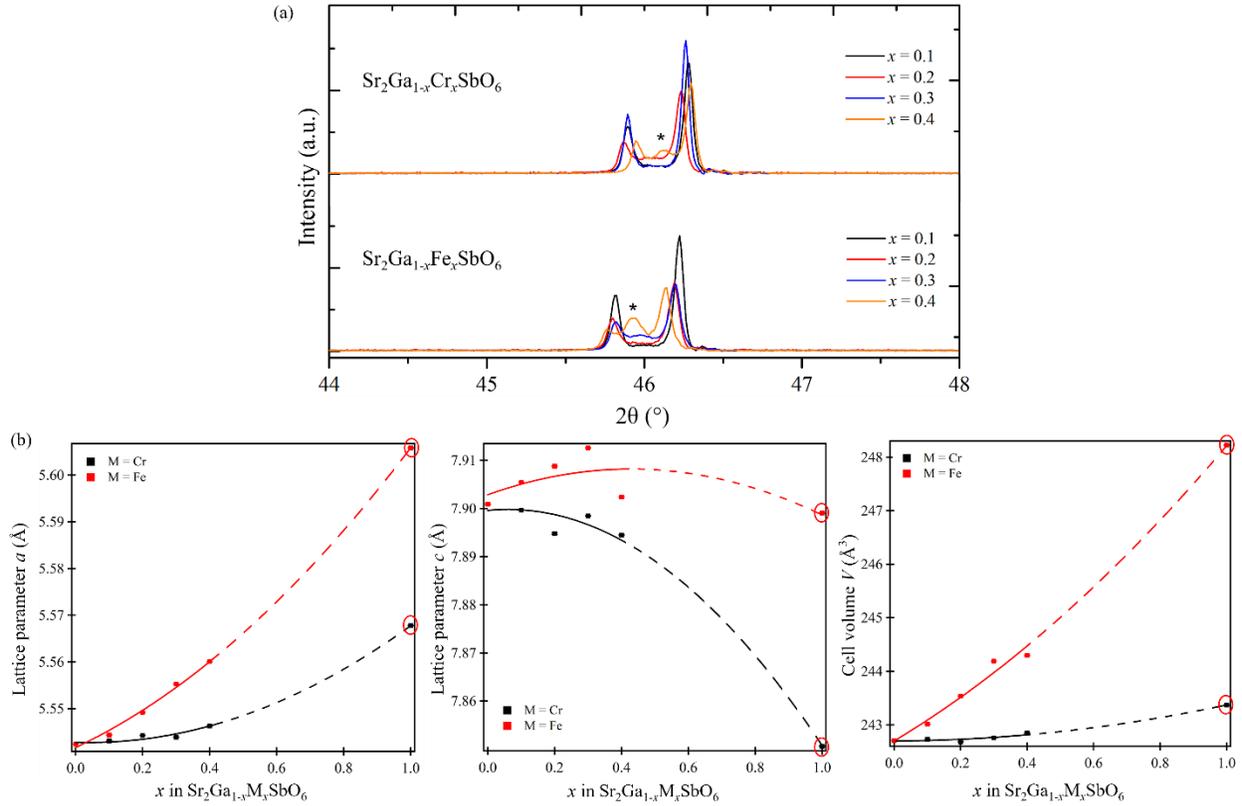

**Figure 2.** (a) Stacked lab X-ray diffraction patterns for all compositions of the $Sr_2Ga_{1-x}M_xSbO_6$ series (M = Cr/Fe); (b) lattice parameter $a$, $c$ and cell volume $V$ (error bars are smaller than the data points) plotted for each composition of $Sr_2Ga_{1-x}M_xSbO_6$ series (M = Cr/Fe). (Data points in red circles are extracted from literature for the end-members $Sr_2CrSbO_6$ and $Sr_2FeSbO_6$, which fall in line with the materials reported here.)

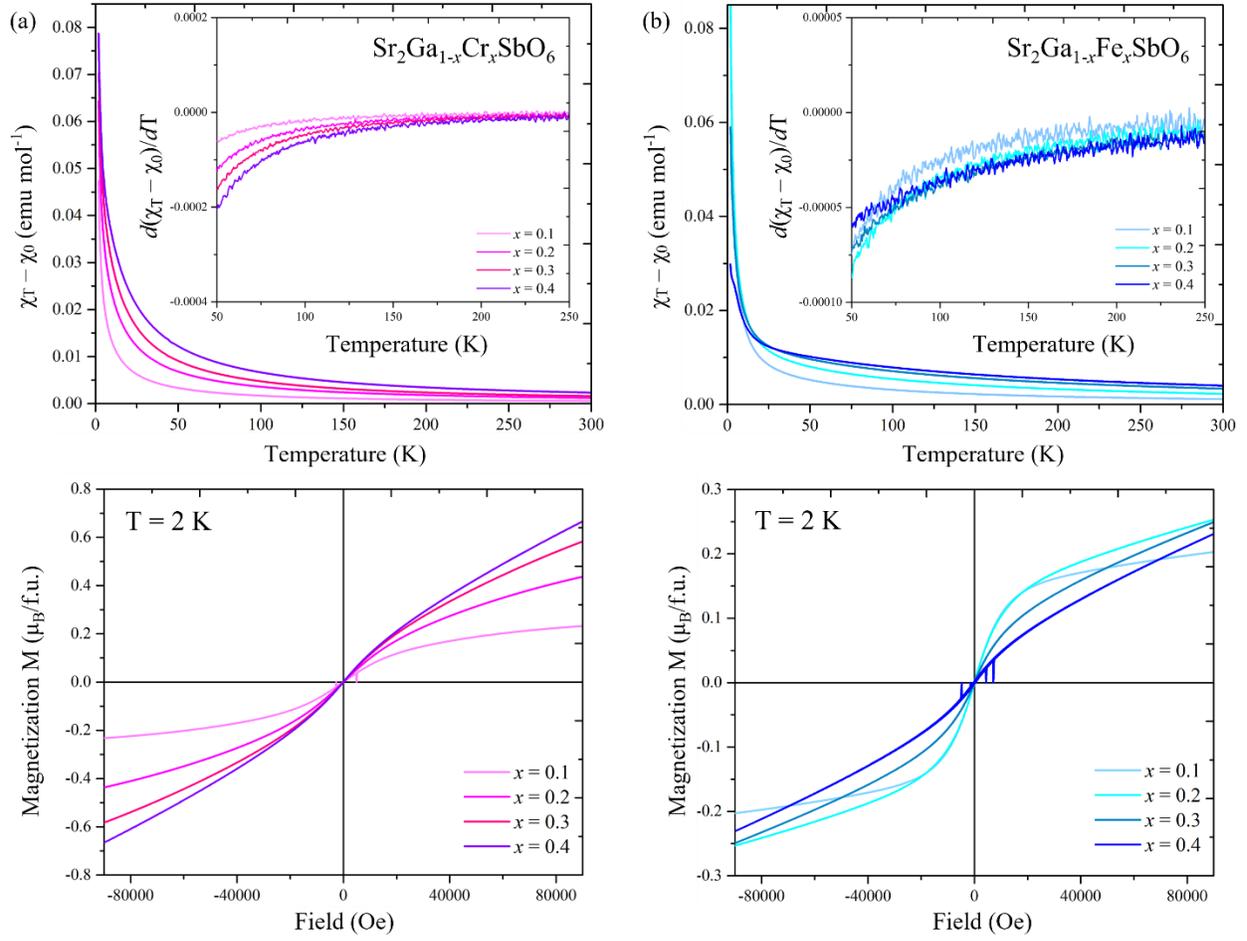

**Figure 3.** The ($\chi_T - \chi_0$) plotted against temperature, with the first derivative plot embedded (top) and the magnetic moment $M$ collected at 2 K plotted against field $H$ (bottom) for each composition in the (a) $Sr_2Ga_{1-x}Cr_xSbO_6$ series and (b) $Sr_2Ga_{1-x}Fe_xSbO_6$ series.

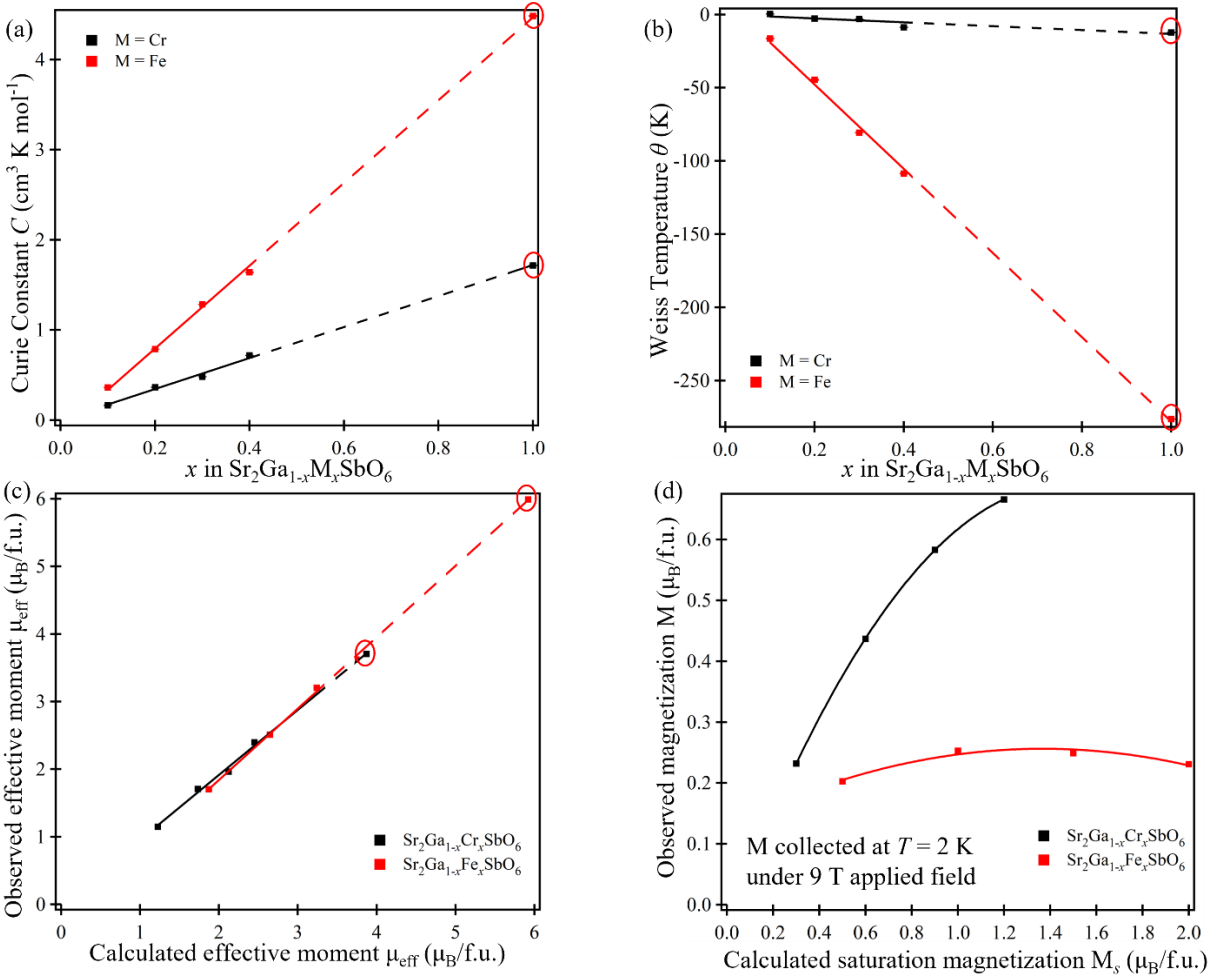

**Figure 4.** (a) The Curie constant and (b) the Weiss temperature extracted from the fitting of paramagnetic susceptibility to the Curie-Weiss law; (c) the observed effective moment per formula unit plotted against the effective moment per formula unit calculated from the spin-only contribution; (d) the observed magnetization at 2 K under an applied field of 9 T plotted against the calculated saturation magnetization (error bars are smaller than the data points) for each composition of $Sr_2Ga_{1-x}M_xSbO_6$ series (M = Cr/Fe). (Data points in red circles are extracted from literature for the end-members $Sr_2CrSbO_6$ and $Sr_2FeSbO_6$, which fall in line with the materials reported here.)

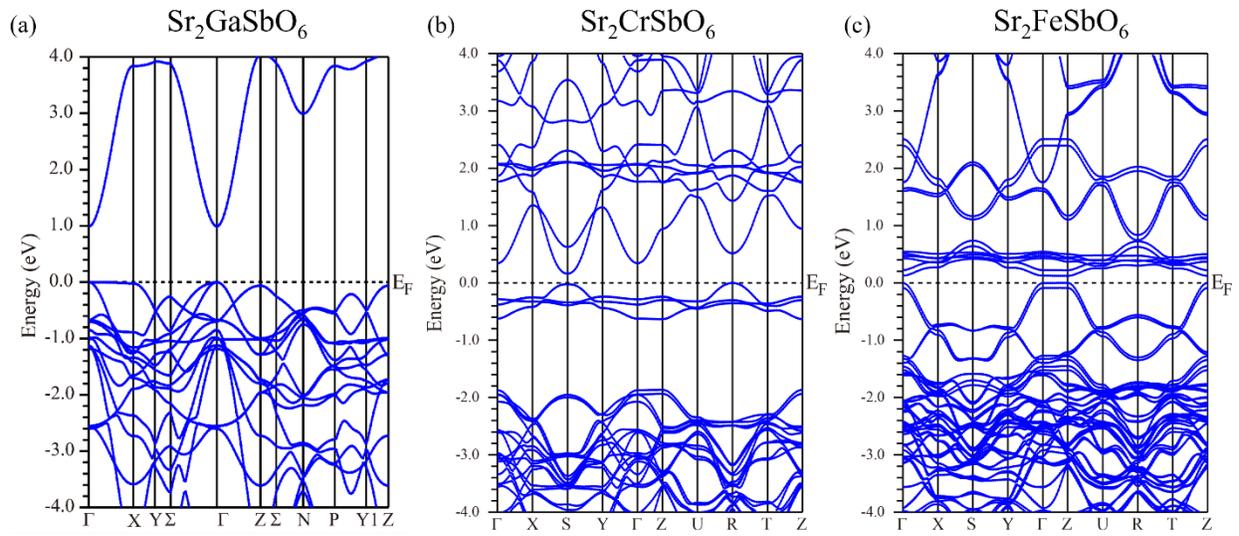

**Figure 5.** Calculated band structures and electronic density of states (DOS) of (a) Sr₂GaSbO₆, (b) Sr₂CrSbO₆ and (c) Sr₂FeSbO₆.

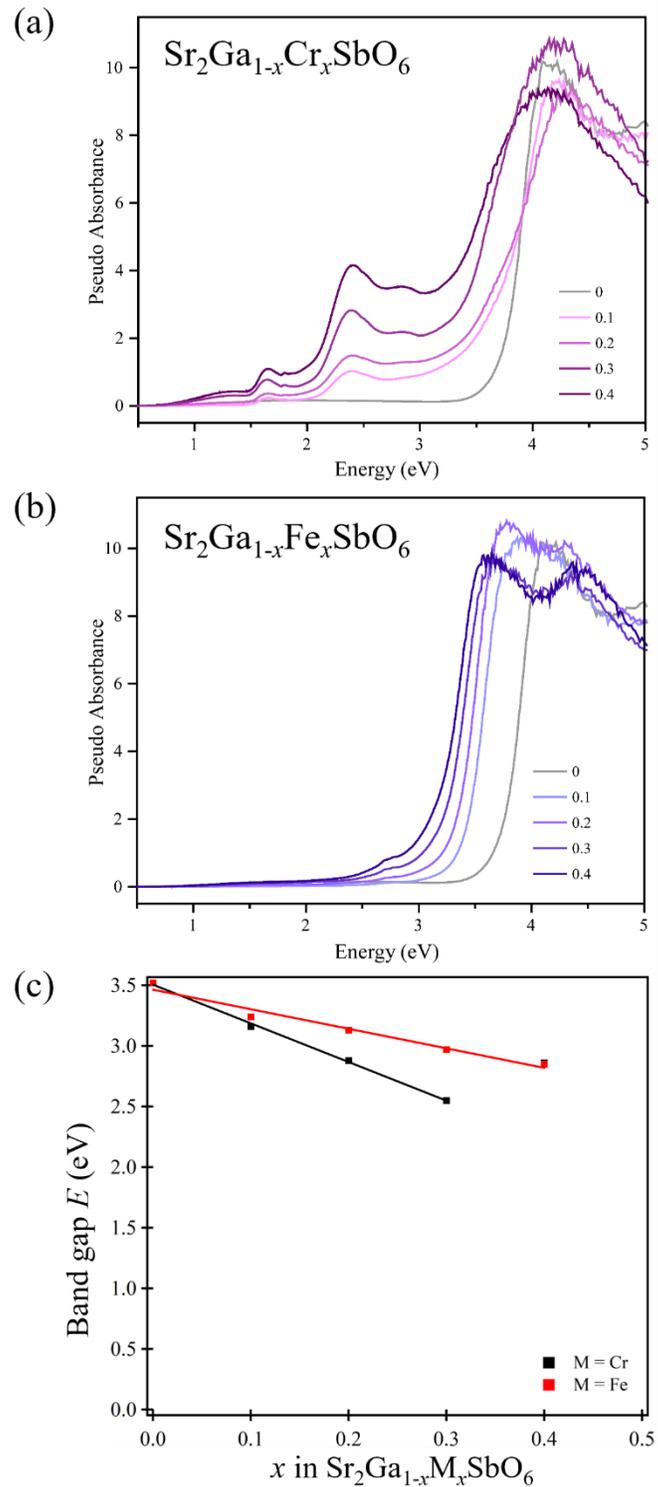

**Figure 6.** The diffuse reflectance spectra of (a) $Sr_2Ga_{1-x}Cr_xSbO_6$ series and (b) $Sr_2Ga_{1-x}Fe_xSbO_6$ series; and (c) the band gaps from Tauc plots obtained by using an indirect transition equation plotted against the doping level $x$ for both Cr- and Fe-doped series.

# References

1   P. Schiffer, A. P. Ramirez, W. Bao and S.-W. Cheong, *Phys. Rev. Lett.*, 1995, **75**, 3336–3339.

2   R. E. Cohen, *Nature*, 1992, **358**, 136.

3   A. H. Macdonald, P. Schiffer and N. Samarth, *Nat. Mater.*, 2005, 4, 195–202.

4   T. Story, R. R. Gafqzka, R. B. Frankel and P. A. Wolff, *Phys. Rev. Lett.*, 1986, **56**, 777–779.

5   T. Dietl and H. Ohno, *MRS Bull.*, 2003, **28**, 714–719.

6   H. Munekata, H. Ohno, ' S Von Molnar, A. Segmuller, L. L. Chang and L. Esaki, *Phys. Rev. Lett.*, 1989, **63**, 1849–1852.

7   T. Dietl, H. Ohno, F. Matsukura, J. Cibert and D. Ferrand, *Science.*, 2000, **287**, 1019–1022.

8   T. Kong, K. Stolze, E. I. Timmons, J. Tao, D. Ni, S. Guo, Z. Yang, R. Prozorov and R. J. Cava, *Adv. Mater*, 2019, **31**, 1808074.

9   T. Dietl, *Semicond. Sci. Technol*, 2002, **17**, 377–392.

10  X. Li and J. Yang, *Natl. Sci. Rev.*, 2016, **3**, 365–381.

11  M.-R. Li, M. Retuerto, Z. Deng, P. W. Stephens, M. Croft, Q. Huang, H. Wu, X. Deng, G. Kotliar, J. Sánchez-Benítez, J. Hadermann, D. Walker and M. Greenblatt, *Angew. Chemie Int. Ed.*, 2015, **54**, 12069–12073.

12  H. Ohno, *Science.*, 1998, **281**, 951–956.

13  J. K. Furdyna, *J. Appl. Phys.*, 1998, **64**, R29.

14  M. A. Peña and J. L. G. Fierro, *Chem. Rev.*, 2001, **101**, 1981–2017.

15  I. Chung, B. Lee, J. He, R. P. H. Chang and M. G. Kanatzidis, *Nature*, 2012, **485**, 486–489.

16  J. Sunarso, S. S. Hashim, N. Zhu and W. Zhou, *Prog. Energy Combust. Sci.*, 2017, **61**, 57–77.

17  M. Retuerto, M. García-Hernández, M. J. Martínez-Lope, M. T. Fernández-Díaz, J. P. Attfield and J. A. Alonso, *J. Mater. Chem.*, 2007, **17**, 3555–3561.

18  E. C. Hunter and P. D. Battle, *J. Solid State Chem.*, 2018, **264**, 48–58.

19  S. Baidya and T. Saha-Dasgupta, *Phys. Rev. B*, 2012, **86**, 024440.

20  A. Faik, J. M. Igartua, M. Gateshki and G. J. Cuello, *J. Solid State Chem.*, 2009, **182**, 1717–1725.

21  P. D. Battle, T. C. Gibb, A. J. Herod and J. P. Hodges, *J. Mater. Chem.*, 1995, **5**, 75–78.

22  N. Kashima, K. Inoue, T. Wada and Y. Yamaguchi, *Appl. Phys. A*, 2002, **74**, 805–807.

23     A. Faik, J. M. Igartua, E. Iturbe-Zabalo and G. J. Cuello, *J. Mol. Struct.*, 2010, **963**, 145–152.

24     H. Rietveld, *J. Appl. Crystallogr.*, 1969, **2**, 65–71.

25     A. Y. Tarasova, L. I. Isaenko, V. G. Kesler, V. M. Pashkov, A. P. Yelisseyev, N. M. Denysyuk and O. Y. Khyzhun, *J. Phys. Chem. Solids*, 2012, **73**, 674–682.

26     P. Blaha, K. Schwarz, G. K. H. Madsen, D. Kvasnicka, J. Luitz, R. Laskowsk, F. Tran, L. Marks and L. Marks, *WIEN2k: An Augmented Plane Wave Plus Local Orbitals Program for Calculating Crystal Properties*, Techn. Universitat, 2019.

27     E. Wimmer, H. Krakauer, M. Weinert and A. J. Freeman, *Phys. Rev. B*, 1981, **24**, 864–875.

28     J. P. Perdew and Y. Wang, *Phys. Rev. B*, 1992, **45**, 13244–13249.

29     R. D. King-Smith and D. Vanderbilt, *Phys. Rev. B*, 1993, **47**, 1651–1654.

30     M. W. Lufaso, R. B. MacQuart, Y. Lee, T. Vogt and H. C. Zur Loye, *J. Phys. Condens. Matter*, 2006, **18**, 8761–8780.

31     S. Berri, *Chinese J. Phys.*, 2017, **55**, 2476–2483.

32     D. L. Wood and J. Tauc, *Phys. Rev. B*, 1972, **5**, 3144.

33     B. J. Wood and R. G. J. Strens, *Mineral. Mag.*, 1979, **43**, 509–518.

34      L. Jin, D. Ni, X. Gui, D. B. Straus, Q. Zhang and R. J. Cava, preprint, arXiv:2109.10405.